\documentclass[aps,twocolumn,groupedaddress]{revtex4}

\usepackage{graphicx}
\usepackage{hyperref}
\usepackage{dcolumn}
\usepackage{bm}
\usepackage{epsfig,amsbsy,bm,amssymb,amsmath}
\usepackage{xcolor} 
\usepackage{tikz}
\usetikzlibrary{shadings}

\newcommand{\hs}{\hspace*}

\newcommand{\np}{\newpage}
\newcommand{\w}{\omega}
\newcommand{\W}{\Omega}

\newcommand{\eref}[1] {(\ref{#1})}
\newcommand{\Eref}[1] {Eq.~(\ref{#1})}
\newcommand{\Fref}[1] {Fig. \ref{#1}}

\newcommand{\ra}{\rangle}
\newcommand{\la}{\langle}
\newcommand{\nn}{\nonumber}
\newcommand{\be}{\begin{equation}}
\newcommand{\ee}{\end{equation}}
\newcommand{\br}{\begin{eqnarray*}}
\newcommand{\er}{\end{eqnarray*}}
\newcommand{\ba}{\begin{eqnarray}}
\newcommand{\ea}{\end{eqnarray}}
\newcommand{\bp}{\begin{minipage}}
\newcommand{\ep}{\end{minipage}}
\newcommand{\bt}{\begin{tabular}}
\newcommand{\et}{\end{tabular}}

\newcommand{\ms}{\vspace*{-5mm}}
\newcommand{\mms}{\vspace*{-2.5mm}}

  \newcommand{\q}{{\bm q}}

 \newcommand{\ig}[1]{\includegraphics[width={#1}]}

\newcommand{\isum}%
{\mathop{\hbox{$\displaystyle\sum\kern-15.2pt\int\kern1.5pt$}}}

\renewcommand{\t}{\tau}

\renewcommand{\q}{\theta}

\newcommand{\ask}[1]{\textcolor{blue}{#1}\hs{-1mm} }
\renewcommand{\ask}[1]{\textcolor{black}{#1}\hs{-1mm} }
\newcommand{\bsk}[1]{\textcolor{blue}{#1}\hs{-1mm} }
\renewcommand{\bsk}[1]{\textcolor{black}{#1}\hs{-1mm} }

\graphicspath{{./}{./PDF/}{./EPS/}{./PNG/}{./Data/}}

\begin{document}
\bibliographystyle{apsrev}

\title{Circular RABBITT goes under threshold: \\\ask{A sensitive probe of
  discrete excitations in noble gas atoms}}

\author{Vladislav V. Serov$^{1}$}
\author{Jia-Bao Ji$^2$}
\author{Meng Han$^{3}$}
\author{Kiyoshi Ueda$^{4,5}$}
\author{Hans Jakob W{\"o}rner$^{2}$}
\author{Anatoli~S. Kheifets$^{6}$}

\affiliation{$^1$Department of Medical Physics, Saratov State University, Saratov
  410012, Russia}
\affiliation{$^2$ Laboratorium f{\"u}r Physikalische Chemie, ETH
  Z{\"u}rich, 8093 Z\"urich, Switzerland}
\affiliation{$^3$ J. R. Macdonald Laboratory, Department of Physics,
  Kansas State University, Manhattan, KS 66506, USA}
\affiliation{$^4$ Department of Chemistry, Tohoku University, Sendai,
  980-8578, Japan}
\affiliation{$^5$ School Physical Science and Technology, ShanghaiTech
  University, Shanghai 201210, China}
\affiliation{$^{6}$Research School of Physics, The Australian
  National University, Canberra ACT 2601, Australia}

 \date{\today}

\begin{abstract}
We introduce circular under-threshold RABBITT (cuRABBITT) as a new interferometric method
to probe discrete electronic excitations in atoms with attosecond resolution. By combining
circularly polarized attosecond pulses with broadband (``rainbow'') spectral analysis, we
directly access two-photon ionization amplitudes and their relative phases. Time-dependent
Schr\"odinger simulations, supported by Green's function theory, reveal strong resonances
in helium and argon and a Cooper-like minimum in xenon. These results demonstrate that
cuRABBITT provides continuous spectral mapping of bound-state resonances and extends
Fano's propensity rule into the under-threshold regime. Our work establishes cuRABBITT as
a powerful attosecond metrology technique, opening the way to polarization-resolved
studies of resonant dynamics in atoms and molecules.

\end{abstract}

\maketitle


Attosecond science provides the shortest controlled bursts of light
currently available, allowing physicists to track the motion of
electrons on their natural timescales ($10^{-18}$ s). These techniques
have transformed our understanding of electron dynamics in atoms,
molecules, and solids, and were recently recognized by the 2023 Nobel
Prize in Physics~\cite{NobelPrizePhysics2023}.  Among the most widely
used methods is Reconstruction of Attosecond Beating By Interference
of Two-photon Transitions (RABBITT) \cite{MullerAPB2002,TomaJPB2002}, which
measures electron wave packet dynamics with sub-femtosecond precision.

In its conventional form, RABBITT employs linearly polarized
extreme-ultraviolet (XUV) and infrared (IR) fields to probe two-photon
ionization pathways.  This approach has enabled studies ranging from
autoionizing resonances and time delays near Fano
resonances~\cite{Gruson2016,Cirelli2018} to spin-orbit and
fine-structure effects~\cite{Turconi2020}.  More recently, circularly
polarized fields have been introduced, unlocking access to dichroic
phases that directly encode partial ionization
amplitudes~\cite{HanMeng2023,Han2023Optica,MengHan2024,Han2025}.
These advances open a new perspective on how angular momentum and
polarization shape ultrafast electron dynamics.

A fundamental modifictation to RABBITT arises when one of the driving
XUV harmonics falls below the ionization threshold. In this so-called
\emph{under-threshold} RABBITT (uRABBITT) regime, the missing
continuum pathway is replaced by excitation through a manifold of
discrete Rydberg
states~\cite{Swoboda2010,Drescher2022,Neoricic2022,Autuori2022,Jiang2025,MengHan2025,Moioli2025}
This situation is not a mere complication: it provides a sensitive
interferometric window onto bound-state excitations and their coupling
to the continuum.  Yet until now, under-threshold effects have not
been systematically explored in circular polarization, where
polarization-resolved observables can expose entirely new selection
rules.

In this Letter, we introduce \emph{circular under-threshold} RABBITT
(cuRABBITT) combined with broadband ``rainbow'' spectral analysis.
This technique allows us to continuously map two-photon ionization
amplitudes and their relative phases across wide energy ranges,
revealing resonances, anti-resonances, and Cooper-like minima in
noble-gas targets.  By benchmarking time-dependent Schrödinger
equation (TDSE) simulations against Green's function theory, we
demonstrate that cuRABBITT not only captures the dynamics of discrete
excitations, but also extends Fano's propensity rule into the
under-threshold regime.  Our results establish cuRABBITT as a general
attosecond metrology tool with polarization control, opening pathways
to disentangle competing resonant channels in atoms and molecules.

\bsk{ While attosecond interferometric techniques based on single
  attosecond pulses and infrared probing have long enabled
  reconstruction of electron wave packets in amplitude and phase
  \cite{NobelPrizePhysics2023,Mauritsson2010}, their observables
  generally represent coherent superpositions of multiple ionization
  pathways and angular-momentum channels
  \cite{Dahlstrom2012,Isinger2019}. As a result, direct access to
  individual two-photon transition amplitudes and their relative
  phases has remained elusive \cite{Kotur2016,Busto2018}, particularly
  in the presence of strong resonant contributions below the
  ionization threshold \cite{Swoboda2010,Neoricic2022}.
Here, we show that combining circularly polarized XUV and IR fields
with under-threshold RABBITT provides a qualitatively new capability:
the dichroic RABBITT phase encodes the ratios and phase differences of
competing two-photon partial-wave amplitudes
\cite{HanMeng2023,kheifets2024characterization}. This enables
angular-momentum-resolved, phase-sensitive spectroscopy of discrete
excitations and allows us to probe the validity and breakdown of
Fano’s propensity rule in a regime that has not previously been
accessible \cite{PhysRevLett.123.133201,PhysRevA.103.L011101}.}

\begin{figure}[h!]
\ig{9cm}{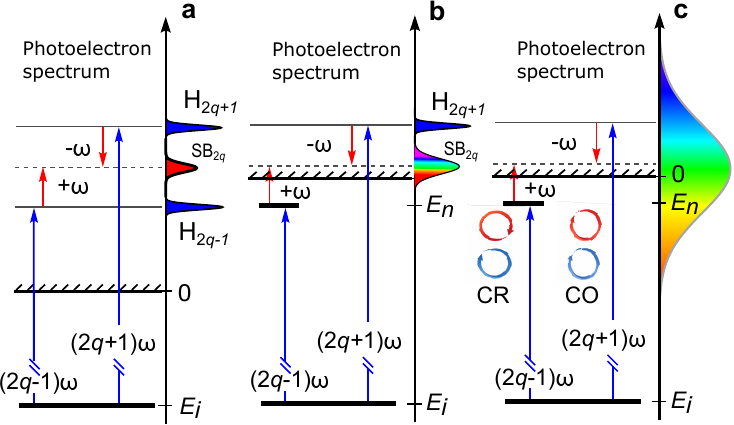}

\caption{ Schematic of RABBITT processes. (a) Standard case: IR
  absorption/emission couples adjacent odd XUV harmonics, forming
  sidebands at $2q\w$. (b) Under-threshold RABBITT (uRABBITT): one
  harmonic lies below threshold and couples via discrete Rydberg
  states. (c) Continuous rainbow RABBITT: a single attosecond pulse
  yields broadband spectra for extended analysis. \ask{The CO and CR cases
  of circular XUV (blue) and IR (red) polarization are illustrated in
  the right panel.}  }
\label{Fig1}
\ms
\end{figure}

Common to all the RABBITT applications is a comb of odd XUV harmonics
$(2q\pm1)\w$ from an attosecond pulse train (APT) which is augmented
by an absorption (marked with $+$ sign in the following) or emission
(marked with $-$ sign) of one driving laser IR photon with the carrier
frequency $\w$.  This IR photon absorption/emission creates sidebands
(SBs) centered at $2q\w$ as illustrated in the photoelectron spectrum
exhibited in \Fref{Fig1}a.  The height of the side\-bands oscillates
at twice the IR photon frequency as the XUV/IR pulse delay $\tau$
varies:
\ms

\be
S_{\rm SB}(\tau) =A+
B\cos[2\omega\tau-C]
\ \ , \ \
C = 2\w\t_a
\ .
\label{RABBITT}
\ee
Here $A,B$ are the RABBITT magnitude parameters and $C$ is its
phase. The latter can be linked with the atomic time delay
$\t_a\simeq \t_W+\t_{cc}$ decomposed into the Wigner time delay $\t_W$ and
the continuum-continuum (CC) component $\t_{cc}$ \cite{Dahlstrom2012}.

If one harmonic energy submerges below the ionization threshold
$(2q-1)\w <|E_i|<(2q+1)\w
\,,
$ 
the corresponding harmonic peak H$_{2q-1}$ disappears from the
photoelectron spectrum. Instead, the missing absorption path of the
conventional RABBITT process can proceed via a series of discrete
Rydberg excitations $E_n<0$. Such an under-threshold or uRABBITT
process is illustrated graphically in \Fref{Fig1}b.  The uRABBITT
process has been observed experimentally in He
\cite{Swoboda2010,Drescher2022,Neoricic2022,Autuori2022,Jiang2025,MengHan2025},
Ne \cite{Moioli2025} and Xe \cite{Villeneuve2017}. Theoretically, it
has also been studied in Ne
\cite{PhysRevA.103.L011101,Kheifets2021Atoms} and Ar
\cite{Kheifets2023}.

With circular radiation, the RABBITT parameters entering
\Eref{RABBITT} become dichroic, i.e. they differ for the co-rotating
(CO) and counter-rotating (CR) XUV and IR fields.  The knowledge of
the dichroic phase $C$ in both cases allows for a retrieval of the
two-photon ionization amplitudes and their phases, not generally
  possible with linear polarization
  \cite{kheifets2024characterization}. More specifically, the
circular XUV photon absorption with $M=1$ drives the initial atomic
state $l_i,m_i\ge l_i-1$ to the uniquely defined intermediate state
with $\ell=l_i+1$. Depending on the CR or CO polarization of the IR
photon, the angular momentum of the final state acquires the two
values $L=\ell\pm1$. The set of the two CO/CR phases allows to
determine the moduli ratio of the two ionization amplitudes and their
relative phase
\cite{kheifets2024characterization,kheifets2025circularly}
\be
\label{ratios}
R^\pm_\ell=\Big|T^\pm_{\ell \to \ell-1}/T^\pm_{\ell \to \ell+1}\Big|
 \ , \  
\Delta\Phi^\pm_\ell=
\arg\Big[T^\pm_{\ell \to \ell-1}/T^\pm_{\ell \to \ell+1}\Big] 
\ee
Here $T^\pm_{\ell \to \ell\pm1}$ are the two-photon ionization
amplitudes stripped of their angular dependence as defined  in
\Eref{LOPT}. 
The moduli ratios $R^\pm_\ell$ are of particular interest because of
the recently formulated Fano's propensity rule in two-photon XUV+IR
ionization processes \cite{PhysRevLett.123.133201}. By virtue of this
rule, the angular momentum is preferably increased or decreased in the
IR photon absorption/emission processes, respectively. This implies
the inequalities $R^+_\ell < 1$ and $R^-_\ell > 1$. These inequalities
have indeed been confirmed by numerical linear
  \cite{Bertolino2020} and circular
  \cite{kheifets2024characterization,kheifets2025circularly} RABBITT
simulations.  Various analytic theories
\cite{PhysRevA.107.043113,PhysRevA.110.013120,Ji2024analytical,PhysRevA.111.043107}
do also generally support the Fano rule in two-photon
ionization. However, at a sufficiently high IR photon frequency, a
large orbital momentum and a low photoelectron energy, a
departure from the Fano rule is predicted
\cite{Ji2024analytical} with the CR polarization being able to produce
$R^+_\ell > 1$. At the photoelectron energies in between the two
regimes of $R^+_\ell < 1$ and $R^+_\ell > 1$, the absorption ratio
passes through a characteristic Cooper-like minimum.

These analytic predictions are hard to verify either numerically or
experimentally because sufficiently low photoelectron energies always
imply the uRABBITT regime. So the circular RABBITT should necessarily
go under threshold. In this Letter, we demonstrate such a circular
under threshold RABBITT (cuRABBITT). In this demonstration we employ a
rainbow spectral analysis which is illustrated graphically in
\Fref{Fig1}b. In the rainbow RABBITT (rRABBITT), each of the dense
grid of energy points in the photoelectron spectrum under the
SB$_{2q}$ is the subject of the time variation \eref{RABBITT} instead
of the overall peak height as in \Fref{Fig1}a.  Such an extended
spectral analysis has proven instrumental to disentangle various
ionization pathways involving auto\-ionizing resonances
\cite{Kotur2016,Gruson734,Busto2018,Isinger2019}, bound
 states \cite{Neoricic2022} and fine-structure splittings
\cite{Turconi2020,Roantree2023}. The same technique is beneficial when
the presence of multiple ionization channels leads to spectral
congestion in atoms \cite{Alexandridi2021} and molecules
\cite{Borras2023}.

The restriction of the rRABBITT is that its span is limited
to the spectral width of the single above-threshold SB$_{2q}$. To span
a sufficiently wide portion of the photo\-electron spectrum, the IR
photon energy $\w$ should be continuously adjusted. This is not
permitted in the present context as the ratios $R^\pm_\ell$ also
change rapidly with $\w$. To circumvent this difficulty, we realize
the continuous rRABBITT which is not limited to any particular
SB. As in an earlier work by \citet{Mauritsson2010}, we
replace a narrow-band APT with a short single attosecond pulse (SAP)
thus producing a broad spectrum overlapping with an extended interval
of the photo\-electron energies as illustrated in \Fref{Fig1}c. This
photo\-electron spectrum is strongly dominated by single XUV photon
ionization. To enhance two-photon ionization and to deduce the
parameters of the cosine $2\w\t$ oscillation in \Eref{RABBITT}, we
subtract the single-photon ionization component from the total
ionization amplitude thus bringing out the net two-photon ionization
contribution. 
%

Our computer simulations have been conducted by solving numerically
the time-dependent Schr\"odinger equation (TDSE) in the single active
electron approximation. The two independently developed computer codes
\cite{PATCHKOVSKII2016153,Serov2024} were used for cross-checking.
%
%
The photoelectron spectrum in the given emission
direction is obtained by using the surface flux method
\cite{TaoNJP2012,0953-4075-49-24-245001,PhysRevA.88.043403}. The
angular and energy resolved RABBITT parameters are deduced by
projecting the time oscillation signal \eref{RABBITT} on the unity,
$\cos2\w \t$ and $\sin2\w \t$ basis.  By defining
  $X=\int_0^{2\pi}S(x)\cos2xdx$ and $Y=\int_0^{2\pi}S(x)\sin2xdx$ with
  $x=\w\t$ we obtain the RABBITT magnitude and phase parameters as
  $\tan^{-1}C=-Y/X$ and $B=2\pi^{-1}\sqrt{X^2+Y^2}$. Meanwhile, the
  same integration of the $\w\t$ time oscillation yields $X=Y=0$ thus
  eliminating the single ionization background.

\begin{widetext}
\begin{figure*}



\ig{0.9\textwidth}{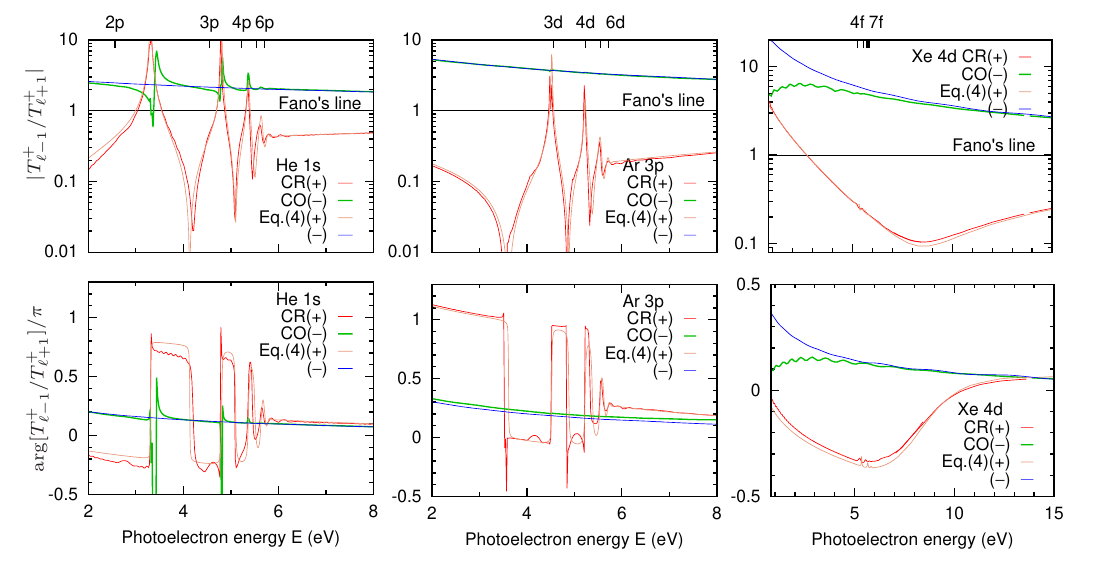}
\mms
\caption{Moduli ratios
  $\left|T^{\pm}_{\ell-1}/T^{\pm}_{\ell+1}\right|$ (top) and phase
  differences and phase differences
  $\arg\!\left[T^{\pm}_{\ell-1}/T^{\pm}_{\ell+1}\right]$ (bottom) for
  He $1s$, Ar $3p$, and Xe $4d$. The crossing of the ``Fano's line''
  ($R=1$) marks deviations from Fano's propensity rule. Resonant
  oscillations appear in He and Ar, while Xe shows a Cooper-like
  minimum with weak resonance structure.
\label{Fig2}
}
\ms
\end{figure*}
\end{widetext}

The amplitude ratios and the phase
differences \eref{ratios} are obtained by fitting the angular
dependent RABBITT phase $C(\q)$ with the following expressions
\cite{kheifets2024characterization}
\ba
\label{angular_sp}
C^{\rm CR/CO}_{l_i=0,m_i=0} &\ell=1\atop =&
 \arg\Big[T_2^{-}T_2^{+*}\Big]
+ \arg\left[P_2(\cos\q)-{T^{\pm}_0 \over T^{\pm}_2}\right]
\\\nn
C_{l_i=1,m_i=0}^{\rm CR/CO} &\ell=2\atop =&
 \arg\Big[T_3^{-}T_3^{+*}\Big]
+ \arg\left[\bar P_3(\cos\q)-{T^{\pm}_1 \over T^{\pm}_3}\right]
\ .
\ea
Here CR/CO orientation corresponds to the +/-- signs and
$
\bar P^1_3 \equiv P^1_3/P^1_1 = \frac32 (-1 + 5\cos^2\theta)
\ .
$
Similar expressions can be derived for higher orbital momenta
\cite{kheifets2025circularly}. 

Results of our numeric simulations are exhibited in \Fref{Fig2} which
displays the moduli ratios $|T^\pm_{\ell-1}/T^\pm_{\ell+1}|$ (the top
row) and the phase differences $\arg[T^\pm_{\ell-1}/T^\pm_{\ell+1}]$
(the bottom row) for He $1s$ (left, $\ell=1$), Ar $3p$ (center,
$\ell=2$) and Xe $4d$ (right, $\ell=3$). Here we choose the laser
photon frequency in the 200~nm spectral range at $\w=6.09$~eV to span
efficiently the whole manifold of the discrete target states.  These
states are revealed in the photo\-electron spectrum at the energies
$E_n+\w$.

As expected from the uRABBITT diagrams of \Fref{Fig1}b and c, it is
the absorption $(+)$ path of the RABBITT process that should probe the
discrete under-threshold excitations most directly. In the circular
RABBITT, the absorption path is encoded into the CR phase. So it is
the complex amplitude ratio $T^+_{\ell-1}/T^+_{\ell+1}$ that should
reveal the resonant structure most clearly and indeed we observe this
structure with the CR orientation in He $1s$ and Ar $3p$. In He, some weaker
resonant structure is also present at the CO $(-)$ orientation.

In the top row of panels, we draw the Fano's line $R=1$ that divides the
$R^{-}>1$ and $R^{+}<1$ ratios provided they comply with the Fano's
propensity rule. The Fano's line is crossed and the rule is
departed in all the considered target atoms. In He, the line is
crossed with both the CR and CO orientations while in Ar and Xe it is
the CR ratio that crosses this line.

The strong resonant behavior seen in the cases of He $1s$ and Ar $3p$
can be interpreted qualitatively within the lowest order perturbation
theory (LOPT). In this framework, the two-photon ionization amplitudes
can be presented as \cite{Dahlstrom2012,Drescher2022}
\ba
\nn
T^{\pm}_{\ell\pm1}(E=k^2/2) &\propto& {1\over i}
{\mathcal E}_\W {\mathcal E}_\w
\left\{
\sum_{E_{n\ell}<0} +\int_0^\infty \hs{-3mm}d\kappa^2
\right\}
(-i)^L e^{i\eta_L} 
\\
&&\hs{-2.8cm}
\times
\left[
{\la kL\|r\|n\ell\ra \la n\ell\|r\|n_il_i\ra
\over E_i+\W^\pm-E_{n\ell}-i\gamma}
+
{\la kL\|r\|\kappa\ell\ra \la \kappa\ell\|r\|n_i l_i\ra
\over E_i+\W^\pm-\kappa^2/2-i\gamma}
\right]
\label{LOPT}
\ea
Here ${\mathcal E}_\W, {\mathcal E}_\w$ are the spectral contents of
the XUV and IR fields, respectively, while $\la n_il_i\|, \la
\kappa\ell\|$ and $\la kL\|$ are the initial, intermediate and final
electron states defined by their linear and angular momenta. The first
term in the second line of \Eref{LOPT} describes the discrete
excitation\-s whereas the second term contains the CC
transitions. The first term becomes singular at the excitation energy
$E_i+\W^+=E_{n\ell}$. The second term remains regular and can be
evaluated analytically
\cite{PhysRevA.107.043113,PhysRevA.110.013120,Ji2024analytical,PhysRevA.111.043107}. As
shown by \citet{Drescher2022}, the singular term manifests itself by
the series of resonances and anti-resonances each accompanied by a
$\pi$ up and down phase jump. It is exactly this behavior that is seen
in the phase diagrams of the bottom row of \Fref{Fig2} in the cases of
He (both the CR and CO) and Ar (CR only). Due to a much larger
threshold energy, the Xe $4d$ ratios remain largely smooth. Here, the
CR ratio displays a deep Cooper-like minimum and crosses the Fano's line
near the threshold as predicted in \cite{Ji2024analytical}. The only
trace of discrete excitations can be observed in very minor
oscillation of the CR phase.

For a more quantitative evaluation of \Eref{LOPT} we adopt the Green's
function technique  e.g ~\cite{Fuda1973,Magrakvelidze2016}. 
The partial-wave Green's function is constructed as
\be
G_\ell(r,r';E) =
W^{-1}[f_\ell,h_\ell]
\, f_\ell(r_<,E)\, h_\ell(r_>,E) ,
\ee
where $r_< = \min(r,r')$, $r_> = \max(r,r')$, and
$W[f_\ell,h_\ell]$ is the Wronskian of the regular $f_\ell$ and the
outgoing wave $h_\ell$ solutions of the radial Schr\"odinger equation with the
short-range and the Coulomb potentials.
As discussed in \cite{Ji2024analytical}, for the XUV-IR two-photon
transition, the regular solution for the intermediate state $f_L(r_< ;
E \mp \omega)$ is mainly contributing to the bound-continuum
transition, while the out-going wave solution $h_L(r_> ; E \mp
\omega)$ mainly contributes to the CC
transition, where $E$ is the final electron kinetic energy, $\omega$
is the photon energy of the dressing field, and $E - \omega$ and $E +
\omega$ are the energies of the intermediate states in the absorption
and emission pathways, respectively.

The regular solution can be numerically obtained using the Numerov's
method \cite{allison1970numerical} with the initial behavior
$rf_\ell(r;E) \propto r^{\ell+1}$ at $r \to 0$. For $r > R_0$, the
regular solution is the linear combination of the regular and
outgoing wave solutions to the hydrogenic problem, both of which are real
functions and can be numerically computed using well-established
algorithms \cite{thompson1985coulcc}. After that, the out\-going
solution is backward propagated using the Numerov's method from $r = R_0$
to $r = 0$. Thus, the two linearly independent solutions to the
Schr\"odinger equation with the Coulomb potential added with a
short-range potential are both numerically obtained.  

In order to extend the numerical evaluation into the case where the
intermediate state of the absorption pathway is under threshold (has a
negative kinetic energy), the potential of the inner region (e.g. $r <
200~\mathrm{a.u.}$) is raised by $\omega$ with a smooth edge (e.g. the
Gauss error function), which creates a wide potential barrier and
converts the Rydberg states lying between $-\omega$ and $0$ into shape
resonances, which can be treated in the same framework as described
above with scattering energy raised by $\omega$ and without further
modification.  In order to simulate the effect of the spectral
broadening due to the finite dressing pulse duration, a complex
potential is used. The inner potential is raised by $\omega -
\mathrm{i}\gamma / 2$ instead of $\omega$, with $\gamma = (n^3 /
n_\mathrm{eff}^3) \gamma_0$, where $n_\mathrm{eff} =
\sqrt[3]{2/\gamma_0}$ is the limiting principal quantum number of the
Rydberg series and  $\gamma_0 = 6 \times 10^{-3}$~a.u. which matches best
with the TDSE calculation.

The Green's function evaluation for the amplitudes \eref{LOPT}
  produces the moduli ratios and the phase differences which agree
  rather closely with the TDSE results for all the targets exhibited
  in \Fref{Fig2}.

\begin{figure}[t!]

\hs{-5mm}
\ig{1.1\columnwidth}{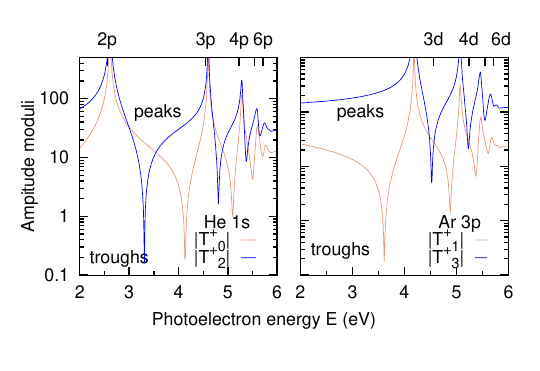}
\ms\ms\mms
\caption{ Individual two-photon amplitudes. Left: $\lvert
  T^{+}_{0}\rvert$ and $\lvert T^{+}_{2}\rvert$ for He $1s$. Right:
  $\lvert T^{+}_{1}\rvert$ and $\lvert T^{+}_{3}\rvert$ for Ar
  $3p$. Resonant peaks align across channels, but displaced
  anti-resonances (troughs) generate the oscillatory ratios of
  \Fref{Fig2}.}
\label{Fig3}
\ms
\end{figure}

By validating the Green's function approach, we can now look at the
amplitudes $T^+_{\ell\to\ell-1}$ and $T^+_{\ell\to\ell+1}$
individually rather than at their ratio which is the only result of
the numerical TDSE simulations. The moduli of the amplitudes
$|T^+_{\ell\pm1}|$ with $\ell=1$ in He $1s$ and $\ell=2$ in Ar $3p$
are displayed in the left and right panels of \Fref{Fig3},
respectively. As expected from \Eref{LOPT}, both pairs of amplitudes
pass through the series of resonances and anti-resonances and display
the set of peaks and troughs. Quite understandably, the resonant peaks
of both the $\ell\to\ell\pm1$ amplitudes match the same set of
discrete energies shifted by the photon energy $\w$. However, quite
remarkably, the anti-resonances and troughs are displaced between
$\ell\pm1$ amplitudes because of the different strength of the
non-resonant continuum. This displacement brings about the strong
oscillatory structure into the amplitude ratio
$T^+_{\ell\to\ell-1}/T^+_{\ell\to\ell+1}$ as displayed in
\Fref{Fig2}. Rather interestingly, the ratio becomes less
oscillatory in the Ar CO case in comparison to He, and almost flattens in
Xe with both the CO and CR orientations. In the latter atom, because of
a considerably larger ionization threshold, the strengths of the two
non-resonant continua equalize and the Cooper minimum further
suppresses the resonant structure.

In conclusion, we have demonstrated that circular under-threshold
RABBITT (cuRABBITT), when combined with rainbow spectral analysis
driven by a single attosecond pulse, provides a uniquely sensitive
probe of discrete excitations in noble gas atoms. 
\ask{
The central advance of this work is the demonstration that circular
under-threshold RABBITT enables direct, phase-resolved access to
individual two-photon transition amplitudes, rather than only to
composite photoelectron wave packets. By exploiting the dichroic phase
difference between co- and counter-rotating ionizing and dressing
fields, we isolate the relative strengths and phases of competing
angular-momentum channels and track their evolution across discrete
bound-state resonances. This capability reveals pronounced resonance-
and anti-resonance-induced violations of Fano's propensity rule and
exposes a Cooper-like minimum in the counter-rotating channel, effects
that are not observable with linearly polarized or non-dichroic
interferometric schemes.}
Our approach enables
continuous mapping of two-photon ionization amplitudes and their
phases across extended photoelectron energy ranges, revealing
resonances and anti-resonances that manifest strongly with
counter-rotating fields and vanish in the co-rotating
configuration. By benchmarking numerical TDSE simulations against
analytic Green's function theory, we resolved not only the resonant
peaks but also the displaced anti-resonances between competing
continua, thereby uncovering the mechanism behind the strong
oscillatory structures in helium and argon and their suppression in xenon.
\ask{Although the currently selected 200~nm spectral range can be
  accessed with frequency-converted femtosecond lasers or modern
  OPCPA-based sources, our results remain valid for the more commonly
  used 800~nm spectral range. The only limitation is that not all
  members of the discrete Rydberg series can be resolved
  simultaneously}

Most importantly, we showed that cuRABBITT extends Fano's propensity
rule into the under-threshold regime, where departures from the
conventional selection rule appear near resonances and at Cooper-like
minima. This establishes a new paradigm in attosecond interferometry,
granting direct access to bound-state excitations and
continuum–continuum coupling with polarization control. The technique
paves the way for future experimental studies that exploit dichroic
phases to disentangle competing pathways in atoms and molecules, and
for theoretical work on generalized propensity rules in complex
multielectron systems.

\paragraph*{Acknowledgment:}  

This work is supported by the Australian Research Council Discovery
Grant DP230101253.  Resources of the National Computational
Infrastructure NCI Australia have been utilized.
J.-B.Ji acknowledges the funding from the ETH grant 41-20-2.
\np


\end{document}